\documentclass{article}

\usepackage{PRIMEarxiv}

\usepackage[utf8]{inputenc} 
\usepackage[T1]{fontenc}    
\usepackage{hyperref}       
\usepackage{url}            
\usepackage{booktabs}       
\usepackage{amsfonts}       
\usepackage{nicefrac}       
\usepackage{microtype}      
\usepackage{lipsum}
\usepackage{fancyhdr}       
\usepackage{graphicx}       
\graphicspath{{media/}}     
\usepackage{natbib}
\usepackage{graphicx}
\graphicspath{ {./} }
\usepackage{multirow}

\title{Fairness Implications of Heterogeneous Treatment Effect Estimation with Machine Learning Methods in Policy-making
}

\author{
  Patrick Rehill \\
  Centre for Social Research and Methods \\
  Australian National University \\
  Canberra\\
  \texttt{patrick.rehill@anu.edu.au} \\
   \And
  Nicholas Biddle \\
  Centre for Social Research and Methods \\
  Australian National University \\
  Canberra\\
  \texttt{nicholas.biddle@anu.edu.au} \\
}

\begin{document}
\maketitle

\begin{abstract}
Causal machine learning methods which flexibly generate heterogeneous treatment effect estimates could be very useful tools for governments trying to make and implement policy. However, as the critical artificial intelligence literature has shown, governments must be very careful of unintended consequences when using machine learning models. One way to try and protect against unintended bad outcomes is with AI Fairness methods which seek to create machine learning models where sensitive variables like race or gender do not influence outcomes. In this paper we argue that standard AI Fairness approaches developed for predictive machine learning are not suitable for all causal machine learning applications because causal machine learning generally (at least so far) uses modelling to inform a human who is the ultimate decision-maker while AI Fairness approaches assume a model that is making decisions directly. We define these scenarios as indirect and direct decision-making respectively and suggest that policy-making is best seen as a joint decision where the causal machine learning model usually only has indirect power. We lay out a definition of fairness for this scenario – a model that provides the information a decision-maker needs to accurately make a value judgement about just policy outcomes – and argue that the complexity of causal machine learning models can make this difficult to achieve. The solution here is not traditional AI Fairness adjustments, but careful modelling and awareness of some of the decision-making biases that these methods might encourage which we describe.
\end{abstract}


\section{Introduction}
Methods to estimate heterogeneous treatment effects (HTEs) with causal machine learning methods promise to give policy-makers access to doubly-robust, non-parametric, individual-level estimates of the impact of a policy \citep{lechner_causal_2023}. This is an intriguing possibility. Data driven exploration of treatment effects may be useful where the peculiarities of a program make existing theory a poor guide to drivers of heterogeneity but researchers have access to a large amount of data \citep{Levin-Rozalis2000Abduction:}. However, there are also potential problems that arise from the complexity of these methods. There is an entire literature of critical AI studies which has documented many cases where governments have misused AI systems (more often out of misunderstanding than a genuine desire to harm some group) leading to poor outcomes \citep{IreniSaban2022Ethical}. While these stories all focus on predictive machine learning (ML) applications (i.e. models focused on predicting outcomes rather than predicting treatment effects \citep{Imbens2015Causal}), one cannot help but worry that those who would use causal ML tools may also stumble into poor outcomes in the absence of a critical AI literature examining these methods specifically. While there are many kinds of poor outcomes – transparency issues, technological failures and privacy concerns to name just a few – this paper will focus on one kind of failure – unfairness in outcomes.

In predictive machine learning applications, it can often be useful to use \textit{AI Fairness} techniques to help make fairer decisions. These approaches broadly try to fit models that equalise outcomes to some extent across different sensitive characteristics like race or gender \citep{Mehrabi2019survey}. They in effect trade-off model fit on (often biased) training data for more equitable outcomes. In general, this involves not just making sure that sensitive characteristics are not included as features when training (so called fairness-through-unawareness), but trying to address the ways these characteristics affect other variables that can proxy these attributes as well \citep{Corbett-Davies2018Measure}. For example, it is not enough to stop a model from penalising an individual directly for being black, these approaches also aim to stop a model from penalising an individual based on living in a majority-black zip code.

Fairness is particularly a topic of debate in policy-making where unfair decisions backed up by the power of the state can be a very dangerous combination \citep{IreniSaban2022Ethical}. However, this paper will argue that existing \textit{AI Fairness} approaches do not for the most part belong in causal machine learning pipelines. The reason for this is that causal machine learning methods exist to help decision-makers understand a data-generating process, not to autonomously make a decision. We call the difference between models that make decisions and models that inform decisions direct and indirect decision-making power. Only models with direct power are suitable for current \textit{AI Fairness} approaches. There are some uses of causal ML models that fall into this category as HTE-estimating models are empowered to make decisions either by directly using the estimates HTEs or by using these estimates as the inputs to a separate policy learning model. However, for the most part, we see causal models as having indirect power and as such as requiring their own set of solutions.

For indirect applications of HTE learners we define an alternative type of unfairness where the complexity of causal ML methods prevents a decision-maker from making an informed value judgement about the consequences of a decision. We see policy-making with these tools as a kind of joint decision-making process where the model informs a human who makes the final decision. One important role of the human here is to weigh fairness concerns that are likely to be more sophisticated than simple mathematical rules. The role of the algorithm is to provide the best possible picture of the data generating process (DGP) to support this decision. This can be seen as a kind of \textit{AI Fairness} focused only on what \citet{Mittelstadt2016ethics} calls epistemic concerns (the evidence provided by analysis) rather than normative concerns (the effect of decisions made with evidence). While all kinds of simpler analytic approaches can also yield misleading evidence, the key difference that makes this an \textit{AI Fairness} issue is the complexity of causal ML models. Essentially, inscrutable models can cause decision-makers to make decisions that are actually worse than the ones they would make with more traditional analysis. We view this as a kind of automation bias \citep{Alon-Barkat2023Human-AI}.

In order to maintain clarity when talking about two kinds of fairness, \textit{AI Fairness} in the formal sense will be italicised while fairness in a more general sense of a value judgement of fairness will be spelled with a not be and generally not capitalised. To restate the contention of this paper in this language, \textit{AI Fairness} approaches can be useless or even harmful when trying to achieve fairness in decisions supported by CML.

The focus of this paper is on government policy-making, but governments are not the only powerful organisations that make decisions that affect people's lives, it is likely that much of this analysis is applicable to other powerful decision-making entities as well. Businesses and not-for-profit organisations also engage in decision-making that might take advantage of causal machine learning algorithms and which could have impacts (positive and negative) that are just as real as decisions made by governments \citep{hunermund_causal_2021}. For example healthcare providers could use CML methods to inform treatment decisions, businesses can use uplift modelling based on these methods for price or service discrimination (in the economic sense of the word 'discrimination') \citep{olaya_survey_2020,chen_causalml_2020}, a large firm could use them to evaluate employee performance in terms of a treatment effect on key metrics. None of this is necessarily good or bad, but it does pose new questions about the morally right and wrong uses of the technology. The main theoretical framework we rely on from \citet{Mittelstadt2016ethics} is sector-agnostic and so while public policy is our area of focus, much of this analysis could apply more broadly to CML-driven decision-making. In fact, it is arguable that while governments have particular powers with regards to regulating individual behaviour, innovative firms --- which lack the expectations of transparency and accountability that come with being a democratic government \citep{IreniSaban2022Ethical} or government's often conservative attitude to adopting novel technologies (at least outside the military) \citep{hinkley_technology_2023} --- are likely to be the first organisations through which the effect of causal machine learning is felt in the real world. Indeed many firms are already embracing the technology \citep{hunermund_causal_2021}.

This paper is divided into 5 sections (including this introduction and a conclusion). Section \ref{intro-cml} provides a brief introduction to the two central ideas of the paper, causal machine learning methods and \textit{AI Fairness}. Section \ref{type-power} defines the direct and indirect power of machine learning tools in government decision-making and discusses how this relates to \textit{AI Fairness} ideas. Section \ref{fair-cml} argues that traditional \textit{AI Fairness} approaches are harmful in the case of causal machine learning systems except in cases where the power of these methods is very direct, such as policy learning. It then lays out what concerns we should have about unfair policy outcomes from causal machine learning and what can be done about them.

\section{Introducing causal machine learning and \textit{AI Fairness}} \label{intro-cml}
\subsection{Causal machine learning}
Causal machine learning is a broad term for three different families of methods which all draw inspiration from the same source, and can be used in tandem for some applications, but are best understood separately. The first is methods used to estimate primarily average treatment effect; the most widely used method for this family is double machine learning \citep{Chernozhukov2018Double/debiased}. It would also cover work on some meta-learners \citep{Künzel2019Metalearners} and older work like causal SVM \citep{Imai2013Estimating}. These methods are focused on estimating ATEs often in high-dimensional, observational data under the assumption that controlling on observables achieves causal identification. 

The second family of methods are focused on estimating heterogeneous treatment effects (HTEs).\footnote{ There are several different types of estimates of HTEs. In practice estimated HTEs (as opposed to unmeasured heterogeneity) are equivalent to conditional average treatment effects (CATEs) (conditioning on a set of covariates). These estimates can yield group average treatment effects (GATEs) or individual treatment effects (ITEs). This paper will use the generic term HTEs for the most part.} A popular method here is the causal forest \citep{Athey2019Generalized,Wager2018Estimation} which uses a random forest made up debiased decision trees to minimise the R-loss objective \citep{Nie2020Quasi-Oracle} in order to estimate HTEs (generally after double machine learning is applied to account for confounding). There are also other approaches such as generic methods with R-Learner \citep{Nie2020Quasi-Oracle,Semenova2021Debiased}, single causal trees \citep{Athey2016Recursive} and causal Bayesian Additive Regression Trees \citep{Hahn2020Bayesian}. Finally, an offshoot of the HTE estimation work is the literature on policy learning. This work uses an ML algorithm to learn rules for policy targeting from heterogeneous treatment effect estimates. Here, the effect estimate does not actually have to be generated by a causal ML method, but in practice it generally is and very much fits into this research program \citep{Amram2020Optimal,Athey2020Policy,McFowland2020Prescriptive,Zhou2018Offline}. These rules can be found under a budget constraint to give rules for allocating a scarce good \citep{Zhou2018Offline}.

The reason for considering these approaches together is that the collective labelling of them as causal machine learning tells us something about how they are likely to be used in practice. They are cutting-edge methods that are relatively new in policy research and so there is not much existing expertise in their use. They present new possibilities in automating the selection of models, removing many of the model-design decisions that a human researcher makes and the assumptions that come with these decisions, but also rely on black-box models in a way traditional explanatory models do not \citep{Breiman2001Statistical}. While there are many different methods that could be discussed in this paper, we will focus on the causal forest from the \textit{grf} package in R \citep{Athey2019Generalized}. This involves a local centring step which essentially adjusts for confounding using double machine learning with two random forest nuisance models (one estimating outcome and one estimating treatment propensity), a causal forest for HTE (and ATE) estimation, and optionally fitting an optimal policy tree based on doubly robust scores for policy learning with the policytree package \citep{Sverdrup2020policytree:}.

\subsection{\textit{AI Fairness} and AI fairness}
\textit{AI Fairness} is a field concerned with the harms that can be done (particularly to already marginalised groups) by employing AI systems to make decisions. There are countless definitions of \textit{AI Fairness} though almost all essentially boil down to the idea that it is not enough to simply omit protected attributes from training an AI model; one must actively protect these attributes by ensuring that discrimination is not occurring through seemingly innocuous variables \citep{Corbett-Davies2018Measure,Mehrabi2019survey}. There are many different ways of achieving this, such as trying to equalise beneficial or adverse outcomes by group or trying to be counterfactually fair at the individual level and there are many metrics by which to measure this fairness \citep{Mehrabi2019survey}. To make this even more complex, many definitions are not mutually satisfiable meaning that meeting one definition can make it impossible to meet another \citep{Hedden2021On}. This paper is not concerned with debating the merits of different approaches to \textit{AI Fairness} though. Instead it deals with two characteristics that are shared by almost all approaches, an interest in formal fairness rules and a concern not primarily with the model itself, but with the implications of the model in the real world. For example, the fact that models consider some outcomes to be adverse and some to be beneficial and define fairness by the real-world distribution of these outcomes.

\textit{AI Fairness} is one approach to addressing fairness problems in applications of machine learning, but there are others as well. For example, literature on explainable and interpretable AI often stresses the importance of understanding a model in order to identify potentially problematic parts of the prediction process \citep{Lipton2018Mythos,Zhou2022Towards}. For the most part though, this literature is driven by discussion of solutions, with less of a focus on defining exactly what problems they solve. There are, for example, serious philosophical questions about how to treat algorithmic decisions that are supported by the power of the state. \citet{Mittelstadt2016ethics} provides a useful framework here defining the potential concerns across a range of steps in the decision-making process. This framework is reproduced in Table \ref{tab:table}, with full explanations available in \citet{Mittelstadt2016ethics} and \citet{IreniSaban2022Ethical} (the former created the framework, the latter applies it more particularly to government). We can see that transparency in models and \textit{AI Fairness} could easily play a role in alleviating some of these concerns, but they cannot address all concerns.

\renewcommand{\arraystretch}{1.3}
\begin{table}[]
\begin{center}
\caption{\label{tab:table}Concerns around algorithmic decision-making from \citet{Mittelstadt2016ethics}}
\begin{tabular}{|l|l|}
\hline
\textbf{Category}                            & \textbf{Concern}                \\ \hline
\multirow{3}{*}{Epistemic concerns} & Inconclusive evidence  \\ \cline{2-2} 
                                    & Inscrutable evidence   \\ \cline{2-2} 
                                    & Misguided evidence     \\ \hline
\multirow{2}{*}{Normative concerns} & Unfair outcomes        \\ \cline{2-2} 
                                    & Transformative effects \\ \hline
Traceability                        & Traceability           \\ \hline
\end{tabular}
\end{center}
\end{table}

This framework provides a more expansive definition of what may make use of algorithmic decision-making in government unfair, recognising that mathematical definitions of \textit{AI Fairness} are not enough and that fairness decisions in government are inherently value-laden. There are six concerns falling into three groups. Epistemic concerns are those where machine learning algorithms create problematic evidence for decisions. Inconclusive evidence means that there are no valid insights that are drawn from data; inscrutable evidence creates opacity where insights from data are not useable because they are not clear enough; misguided evidence leads to biased insights. Normative concerns relate to the actual effect of decisions made on the basis of algorithmic evidence. Unfair outcomes are cases where systems lead to bias in the outcomes of systems (as opposed to just their outputs). This is the main focus of this paper. Transformative effects concern problems of less privacy and autonomy for individuals where algorithms are given more power over people’s lives. Traceability concerns the ability to hold people (and systems) accountable for the outcomes of algorithmic systems.

What is particularly useful for this work is the split between epistemic and normative concerns. The use of machine learning methods in government is always a joint human-machine decision-making process as at the very least, a human decision-maker is choosing to delegate decisions for which they are ultimately responsible to an algorithm \citep{Citron2007Technological,Edwards2017Slave}. Because of this, the epistemic and normative split is important for understanding fairness. The human decision-maker is entirely capable of stepping in between the epistemic and normative concerns to make the decision themselves based on values and a wide range of evidence thus reducing any normative concerns from an algorithm making decisions in its own right based on a narrow decision criterion. However, where there are still epistemic concerns, it is possible that models can still effect normative concerns even through a human decision maker \citep{Busuioc2021Accountable,Mittelstadt2016ethics}. This indirect effect is the reason why it is still necessary to discuss fairness even for causal models with no direct decision-making power. While traceability concerns are also important, these are outside the scope of this paper given it is focused on fairness rather than transparency and accountability.

\section{Types of power that AI systems have and their implications for fairness} \label{type-power}
\subsection{Direct vs indirect power of AI systems}
What fundamentally separates causal and predictive machine learning in most of their uses for the purposes of fairness is not the specifics of the methods, but the degree to which they are empowered in practice. The key difference here is whether the systems themselves are empowered to make decisions or whether they support a human decision-maker i.e. the extent to which they are free to effect Mittelstadt et al.’s normative concerns or whether a human is intervening. The vast majority of predictive applications can be seen as taking a lot of information from a data-generating process and boiling this down to a decision. Causal models on the other hand are about trying to understand the DGP in as much complexity as is valuable. For this reason, the two applications can be seen as having different types of power. The predictive application wields a kind of direct power; it is used where a decision needs to be made and it makes that decision in a way that is usually inscrutable, but this does not matter since after all the goal is to cut through complexity. The goal of causal systems on the other hand is to try and provide some insight into the complexity of the DGP such that decisions can be made on this basis. This does not mean that the causal system lacks power in the decision-making process though. The kinds of information with which a decision-maker is presented shape their decisions, however, in this case the system’s ability to affect the world is mediated through a human decision-maker \citep{Busuioc2021Accountable,Citron2007Technological}. This becomes a crucial point when discussing \textit{AI Fairness} because the agent here is ultimately a human, not an AI system.

\begin{figure}[!h]
    \caption{The Policy Cycle from \citet{Althaus2018Australian}}
    \centering
    \includegraphics[scale=0.3]{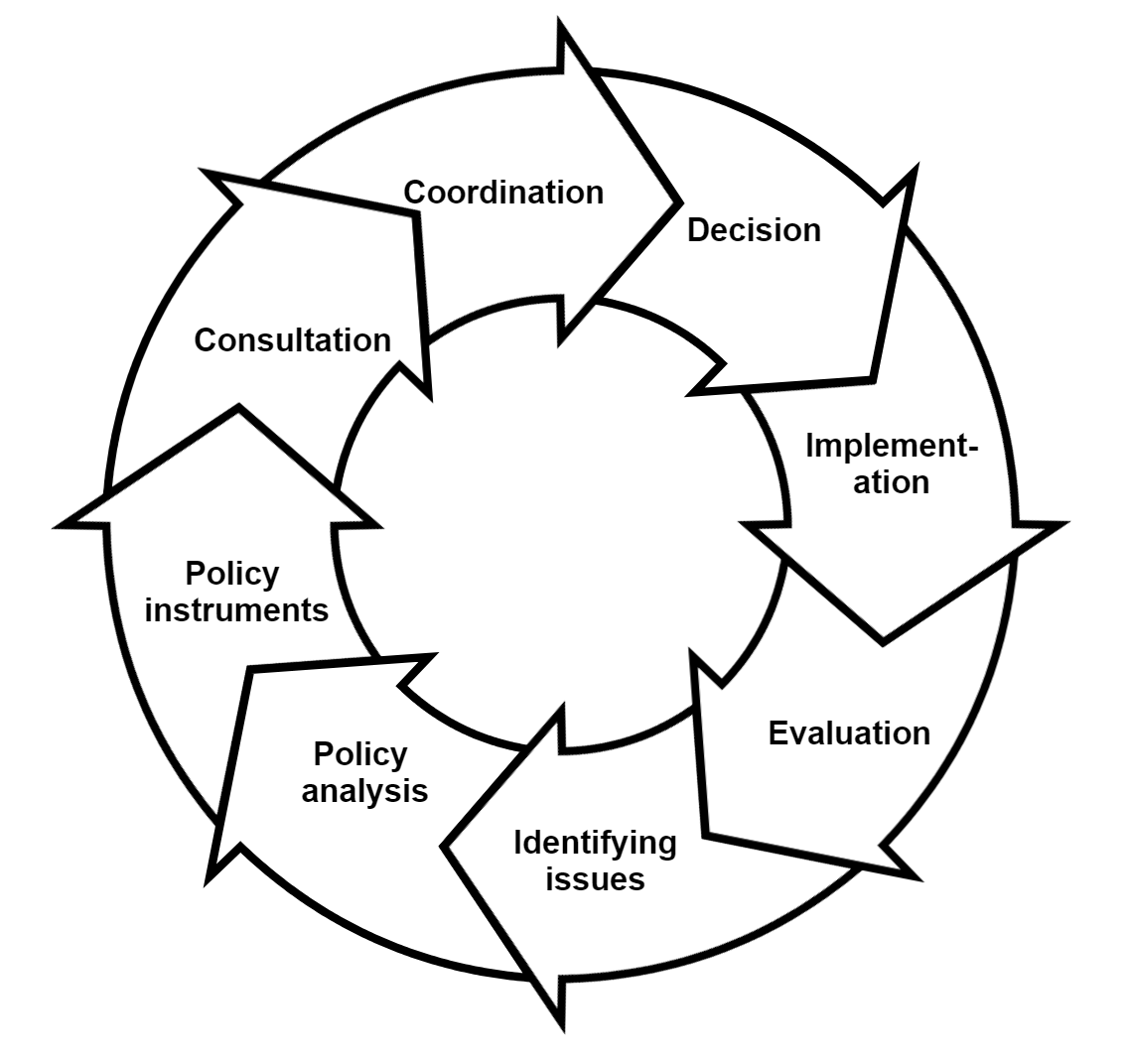}
    \label{fig:fig1}
\end{figure}

It is worth making a brief digression to introduce where evaluation sits in the public policy process for the reader who is unfamiliar with this as this helps to explain what makes indirect power indirect. \citet{Althaus2018Australian} situates evaluation as part of a policy cycle model show in Figure \ref{fig:fig1}. This model was designed to lay out the Australian process, but it is broadly applicable to other jurisdictions and there are many minor variations on a policy cycle model around the world.\footnote{ The policy cycle is an idealised model though. In practice, public policy is generally very under-evaluated (for example, a UK National Audit Office report showed just 8\% of major spending initiatives by the UK Government had robust evaluation plans \citep{NationalAuditOffice2021Evaluating}), it is certainly not the case that most policies are constantly in a state of reassessment. However, the Cycle gives a sense of how evaluation would ideally be used.} The key point is that evaluation is far removed from an actual decision, it helps to identify issues but it is not automatically feeding back into decisions. The five steps between evaluation and decision can be seen as the factors removing a causal ML model from direct power and so when we discuss indirect power, this process should be kept in mind (for more detail on these steps see \citealt{Althaus2018Australian}). These steps consider the possible ways of addressing the problem, build support to help a solution succeed in the real world but they also importantly consider the important value judgements that are crucial to good policy-making.

In situations with direct power an AI model bypasses or incorporates these steps into its algorithm. Coordination and consultation for example are probably going to be bypassed because those who set up the system in the first place decided this was a useful expedient. One could say that a simplified version of identifying issues, policy analysis, policy instruments consideration and decision take place but within the computational logic of the system, generally within tight constraints placed upon it by its designers and the limits of the technology. For example, a mutli-arm causal forest \citep{Athey2019Generalized} could be used to pick one of a menu of policy options, but could not consider novel solutions or evidence beyond a treatment effect estimate.

Returning to the topic of fairness, indirect power lacks some of the fairness problems of direct power, but it brings with it some of its own. Indirect power is still power. It is useful to look at three kinds of poor fairness outcomes that could eventuate from a joint decision between a human decision-maker and a causal model. The first is one where outcomes are poor because of poor modelling (for example, the effect is incorrectly identified based on observational data). Here, the complexity and lack of expertise in using CML models might have played a part as understanding whether an effect is identified in a high dimensional dataset using machine learning methods is ultimately an educated guess \citep{Chernozhukov2018Double/debiased,Hünermund2021Double}, however the remedy here is simply to do better causal modelling, that is to say, it is the modeller who is at fault. In the second case, the model biases a human towards an unfair outcome because of its structure. In the final case, the model works as well as could be hoped, providing an excellent picture of the DGP and the decision-maker correctly accounts for bias in the model, however still proceeds to make a decision that is unfair in some sense, perhaps even un\textit{Fair} by an \textit{AI Fairness} metric. However, here, we suggest that the question of what to do about unfair decisions made by a human decision-maker is well outside the scope of this paper and this field. While it is regrettable, it is not the place of a regression method to try and second-guess the decisions that will be made by those that will employ it as it is ultimately the human that is empowered to make value judgements \citep{IreniSaban2022Ethical}. In all three cases, \textit{AI Fairness} solutions would not be suitable because while there is an AI system in place, its power is indirect and so the solution is to try and make the model give a better picture of the DGP, rather than to equalise predictions across sensitive variables.

One cannot simply say that the difference between causal and predictive models is that one gives AI systems direct power and one gives them indirect power. There are many examples where predictive models are used to inform human decision-making (although this is often misguided and problematic as the systems are inscrutable) \citep{Alikhademi2022review,Angwin2016Machine}. There are also situations where causal models could be given direct power over decisions, policy-learning applications and even cases where a decision-maker defers so much to an HTE estimate that they in practice give it active power over decision-making, though to the best of our knowledge, this has not yet been done in government. These cases show how fine the line can be between direct and indirect power as the former may result not from officially delegated power, but from human decision-makers deferring enough to estimates that the models functionally begin to drive policy decisions, which then have direct impact on individuals. In most cases though implicit or accidental imbuing of causal models with direct power is probably the result of poor use of the methods, meaning that the solution here is improving the ability of the human decision-maker and analyst to support their decision-making. In cases where decisions are being intentionally delegated to an HTE-based model, \textit{AI Fairness} may be useful.

\subsection{What value does \textit{AI Fairness} have?}
For the most part, \textit{AI Fairness} simply is not suitable for HTE estimation, in fact the idea of trying to equalise estimates based on drivers of difference in estimates is anathema to the use of HTE estimating models. The fairness method aims to equalise estimates between groups and the HTE estimating model aims to uncover these \citep{Athey2019Estimating}. If there are differences in HTE estimates by sensitive characteristics, this is exactly the kind of information that might be useful. For example, if a linguistic or ethnic minority is benefitting less from an intervention than the majority, this would be very important to know as it would indicate the program is failing this marginalised group and needs to be changed to be more equitable. Perhaps for example services are not being offered in a language they can understand which is a barrier to uptake of an intervention. In fact, the very kinds of attributes that we would consider sensitive in an \textit{AI Fairness} approach are those we are most likely to be interested in as drivers of heterogeneity. These are likely to be entrenched drivers of disadvantage in society.

There are two clear cases of direct power being given to causal models and therefore \textit{AI Fairness} becomes useful; that of policy-learning and that where a human decision-maker defers completely to an HTE estimating model to the extent that it becomes in practice a very complex policy-learning model. Here the model is actually making decisions about policy eligibility and so its power becomes more direct (although just like predictive systems used by government, as this power is delegated it is not in practice fully direct), it becomes directly responsible for normative concerns. Here the purpose of the model is to make decisions based solely on HTE estimates and so \textit{AI Fairness} approaches become useful again. For instance, to return to our example of a program failing a linguistic minority due to a language barrier, low HTE estimates no longer mean the program has to be fixed to better cater to this group’s needs, instead they would mean making the group ineligible for the service. This is the kind of adverse outcome that \textit{AI Fairness} can help to protect from. For example, \textit{AI Fairness} approaches can ensure equality across groups in policy eligibility even if there is some kind of conditioning based on other variables \citep{Mehrabi2019survey}.

\section{What does it mean for Causal Machine Learning to be fair?} \label{fair-cml}
This section aims to define a new conception of fairness that is analogous to \textit{AI Fairness} for direct power but suited to causal machine learning where models have indirect power. Here we must try to understand not whether the model output itself is unfair, but whether the outcome of the model in the form of a policy decision is an unfair one. In other words, does the model act on the decision-maker to create an unfair outcome; does it effect normative concerns via epistemic concerns? This raises the question of what exactly an unfair outcome is. This is difficult to define mathematically in the way that predictive fairness tends to define the concept as there are no clearly adverse outcomes. \citet{Mittelstadt2016ethics} provides a good answer to this problem. Arguably the greatest ability a human decision-maker has is to weight competing interests and decide what the fair solution to a problem is. This means fairness is simply the outcome that best matches the decision-maker’s value judgement as to what is fair. This helps us for example to differentiate the case that it is perfectly valid to differentiate outcomes on sensitive variables (for example in positive discrimination policies) from other cases where discrimination would not be fair. Instead, we consider unfairness to result from cases where the use of an HTE estimating model creates epistemic concerns that obscure the DGP and therefore, the decision-maker cannot make policies that best fit their value judgement as to what is fair. In this way, the fairness is a subjective thing for the decision-maker to define. This does not mean all definitions are equally correct, rather that it is beyond the scope of the machine learning methods to make this judgement and potentially usurp control over outcomes from the human who is the senior partner in the joint decision-making arrangement. 

What is it though that makes the causal machine learning methods here any different from parametric regression methods? Do the epistemic concerns essentially just boil down to poor causal identification and estimation? Should we consider for example misspecification of a linear regression model for causal inference as an issue of AI fairness as well? While poor causal estimates in any model could be considered a kind of fairness issue under a maximalist definition, the \textit{AI Fairness} lens is not necessarily a useful one in traditional models. The inscrutability of machine learning methods makes it particularly useful to analyse these problems as critical AI problems. Just as how a black-box predictive model creates fairness issues by proxying sensitive variables in a way that the engineers who created it did not intend (and would fix if they knew about the problem and had the tools to do so), the problems of causal fairness also result from complex models that may act in ways its creators did not intend. Our contention is that HTE estimating models can result in two closely linked phenomena automation bias and responsibility gaps. Automation bias is a bias in which humans overly defer to a technological tool; more specifically they do so because their lack of understanding of the inner workings of the tool gives them a false confidence in the tool’s recommendations \citep{Busuioc2021Accountable}. This is a well-established phenomenon in applications relying on predictive machine learning models \citep{Skitka1999Does} and it is easy to see how this might apply across to causal machine learning tools. Responsibility gaps are a result of automation bias where the complexity of new methods and a lack of expertise in their use results in a gap between the effects of an automated system and what is reasonably knowable about the behaviour of that system \citep{SantonideSio2021Four}. While this is generally a concept around accountability for bad outcomes ex post, this gap can be seen as a source of poor outcomes in decision-making as the human decision-making essentially cannot critique the model beyond the point of their understanding and so \textit{de facto} must defer to the system’s judgement. 

Recall the three cases where indirectly powerful systems lead to unfair outcomes from Section 3: poor causal modelling, shaping decisions because of the structure of the model, and unfair decisions made intentionally by a decision-maker. Putting the third case aside as it is – as already discussed – outside the scope of this paper, we see two situations in which indirect power of models can cause unfair outcomes via epistemic concerns that cannot be diagnosed because of the black box nature of the model. The rest of this section discusses how automation bias with HTE-estimating CML methods can result in poor causal identification and in models that emphasise differences in treatment effects which may lead to unfair conditioning of programs where they may not have been conditionalized if other methods were used for evaluation. It discusses how these problems can be remedied; in both cases this means ensuring the humans using such models for decisions and those training causal machine learning models understand the limitations of their models, can challenge models when needed and know when it may be better to employ other possibly less sophisticated methods. While these are not the only possible fairness issues posed by the causal forest, they are the ones that seem obvious even with the causal forest not yet being widely used in policy-making.

\subsection{Convenient identification of causal effects high-dimensional data may encourage lazy identification strategies}
Causal machine learning clearly expands the realm of research by allowing for better modelling-out of biasing effects in high dimensional observational data with double machine learning (either as a stand-alone ATE estimating method or as a local centring step in the causal forest). However, it may also expand the realm of research a user thinks they can or should be able to do even further beyond this. The boundaries of causal machine learning have not yet been tested and codified by researchers and until this happens, it is difficult to tell how far we can trust these (statistically brilliant) methods when their misuse could introduce bias into estimates \citep{Hünermund2021Double}. For the causal forest, this a particularly serious issue as local centring (the causal forest DML orthogonalization stage) is best practice even if experimental or quasi-experimental identification is possible \citep{Athey2019Generalized,Dandl2022What}. For example, the instrumental forest still benefits from local-centring even when there is a strong instrument to help identify effects for reasons that are still not entirely clear \citep{Dandl2022What}. However, even if there is a robust instrument or fully randomised treatment assignment, the use of local-centring could still bias estimates if the model is poorly specified, although the robustness of a model with good identification by quasi-experimental means and poor identification through control-on-observables is yet to be tested.

There is also the specific issue of the presence of post-treatment variables (colliders and mediators) and M structures \citep{Pearl2009Causality:} which could bias analysis just by their inclusion in the dataset while improving the fit of nuisance models \citep{Hünermund2021Double}. It is currently difficult to know just how problematic these factors may be and there are debates about just how much bias they might create in the real world \citep{Ding2015To,Greenland2003Quantifying,Groenwold2021To,Liu2012Implications,Sackett1979Bias}. The problem is that the mystique of double machine learning as a novel method and the opacity of the models that tend to be used in it may stop researchers from realising their estimates suffer from adjusting for post-treatment variables. Some studies that would have used quasi-experimental or experimental methods may have instead opted to use an observational causal ML approach and led to worse policy outcomes. In some cases, not having a study is better than having a biased result as decision-makers can understand the limits of their knowledge and if needed turn to alternative approaches to evaluating a policy (for example participant interviews). It is possible to avoid this issue by ensuring all variables are pre-treatment. This is in fact a requirement for all analysis under the potential outcomes framework \citep{Imbens2015Causal}, however in practice, particularly with a high dimensional dataset not collected for the purposes of causal research, it may be difficult to identify which variables are and are not pre-treatment variables. Pre-treatment variables also do not help address M-bias should it be present.

As an aside, it is also worth noting that the identification of causal effects is only as good as the nuisance models that partial out bias \citep{Chernozhukov2018Double/debiased,Schacht2022So}. Fitting these models requires enough machine learning skill to model this bias, however, this is not a skillset that those who do explanatory modelling in government would generally possess as using machine learning in this way is novel \citep{Breiman2001Statistical,Imbens2021Breiman's}. The gap in competency between being able to fit a supervised machine learning model and being able to fit one well can be large. This means without adequate expertise or even an understanding of the limits of their own skills, analysts might fail to fit models that can partial out bias even if another set of nuisance models could do this well \citep{Schacht2022So}.

Assuming observational causal machine learning lacks internal validity when compared to for example an experimental approach, the real crux of whether it causes harm or not is whether there is a ‘substitution effect’ where the ease of machine learning studies causes a shift from more rigorous to less rigorous methods (including from qualitative designs to treatment effect estimation). If researchers have a very good sense of the method’s strengths and weaknesses and only employ it in cases where it is genuinely the best option available or they employ it in a secondary role like in a sensitivity analysis ala \citet{Chernozhukov2021Causal}, it could provide real value. If on the other hand these methods are used as a lazy way of avoiding making model specification assumptions or finding quasi-experimental / experimental data and give decision-makers biased information which they believe is unbiased due to a lack of expertise in the methods, it will do more harm than good.

\subsection{Emphasising heterogeneous treatment effects could push decision-makers to conditionalize programs}
A second area where HTE estimating methods might bias thinking is in reframing research around questions of conditionalizing policies. With methods that allow for the estimation of HTEs down to the individual level with a large number of covariates, it is possible that it becomes more attractive to conditionalize interventions. This is not necessarily a bad thing; giving policy-makers new tools to understand the benefits of a program can be hugely beneficial and the idea that policy-makers should not be given information if it is available is an alarming one. In fact, this critique is particularly difficult to make given we have deferred to the policy-maker completely on matters of judging the fairness of policies. However, it is worth understanding the implications of HTE and optimal policy learning methods being used in policy research as they may encourage harmful outcomes, if one views programs becoming increasingly conditional on the margin as a harm. The HTE estimates may frame the decision in such a way as to encourage conditionalising when the decision-maker would not do so if they had perfect knowledge of the DGP. In effect, the model is creating misguided evidence the \citet{Mittelstadt2016ethics} definition that may give a false sense of the significance of differences in estimates and disregards arguments for universality of policies, effectively causing a transformative effect where government support becomes more conditional due to access to this new technology and new information on conditional treatment effects. There are two issues here: the explicit focus on conditionalizing programs which has harms not considered by the algorithm which will be addressed in Section \ref{harms-conditionalising}, and the use of HTE as a metric of whether one deserves an intervention or not which will be addressed in Section \ref{deserving}.
\subsubsection{Conditionalising can have harms outside the scope of HTE measurement} \label{harms-conditionalising}
Focusing on HTEs in order to promote conditionalising is not necessarily a bad thing. Giving policy-makers new tools to understand the benefits of a program can be beneficial and the idea that policy-makers should not be given information is an epistemic concern itself. However, it is worth understanding the implications of HTE and optimal policy learning methods being used in policy research as they may frame decisions in ways that encourage harmful outcomes.

Before discussing the harms it is worth considering whether these methods would actually emphasise conditionalising policies. Obviously, the case for such an effect is clearer in cases of policy learning where algorithms have direct power over eligibility. However, this section also contends that it is reasonable to hypothesise that HTE estimation can work to encourage conditioning via indirect power as well. Unfortunately, at this stage there is no evidence either way about whether these tools lead to a specific harm in practice given the immaturity of the critical literature on these methods. However, the literature on joint-decisions with predictive models shows that individuals tend to give algorithms a lot of power to frame their decisions \citep{Busuioc2021Accountable,Citron2007Technological}, meaning it is reasonable to be concerned that a method emphasising differential effects might steer decision-makers towards conditioning programs even if these differences are practically insignificant. This is particularly worrisome given the failure of the causal forest to properly quantify error as it does not account for error in the local centring models \citep{Athey2019Generalized}. This means the methods overstate their own ability to make fine-grained distinctions between treatment effects. Essentially, while it is easy to estimate error in the final model, this unquantified error inherited from the nuisance models could lead to Type I and Type II errors in inference.
 
It does not have to be that the actual estimates themselves are incorrect though. Correct estimates showing a small amount of variation in treatment effects could still be used to justify conditional policies where it is not worth doing so. We can also say that were a government looking to conditionalise a program, HTE estimation and policy learning could make this task easier. These estimates could be useful in conditionalising policy interventions based on the decisions of human decision-makers in ways that standard evaluation methods focused on the ATE or a few drivers of heterogeneity simply are not. We already see for example, some studies using causal forest models to inform conditionalising ‘nudge’ interventions where those with low HTEs are not given a treatment that might have a small positive effect for them \citep{Knittel2021Machine,Murakami2022HETEROGENEOUS}.
 
To the extent that these methods frame policy decisions as problems of optimising eligibility, particularly those under a budget constraint, they are not ideologically neutral. They encourage a certain way of thinking about policies, focused not on actual needs, but on prioritising recipients to keep costs down even if average costs are well below average benefits \citep{Besley1990Means}.\footnote{ While this section talks in terms of cutting eligibility for a program, it is also possible to imagine other policy decisions where governments are trying to maximise the in-scope population; for example auditing or compliance. These different incentives substantively change the fairness implications of the use of HTE estimation for targeting. It is possible to imagine ways in which HTE estimation could change targeting in unfair ways, but because these kinds of decisions are already conditional (because it is presumably not feasible to target the entire population) we cannot say that the availability of HTE estimating methods promotes conditionalization in the way we argue it does for other decisions in this section.} If the time comes to make marginal cuts to a policy, it is easy to use HTE data to find where on the margin the program can be cut most easily even if marginal analysis does not capture the full benefits of the program. There is not necessarily a problem with prioritising efficiency in allocating goods and using a benefit calculation under a budget constraint to make these decisions, but for some policy areas, the best case for universalism is not at all concerned with marginal benefit. Instead, it suggests that there are costs and benefits (e.g. program buy-in, administrative costs, moral imperatives, de-stigmatising participation) that transcend a simple optimisation problem under a constraint \citep{Devereux2021Targeting}. Of course, there are cases in which goods that a government is distributing are genuinely scarce (think for example of governments allocating scarce COVID-19 vaccines when they had been newly approved), but in most cases, a budget constraint for a policy intervention will be a result of earlier decisions around resourcing rather than an actual hard limit (for example, governments can reallocate spending within or between different departments or borrow to finance spending on a program).

\subsubsection{HTE estimates do not define whether someone ‘deserves’ a treatment} \label{deserving}
By estimating HTEs and using them in decision-making, the HTE may become a de facto measure of whether one gets a policy intervention or not and therefore whether one ‘deserves’ an intervention or not. This is a conflation of a purely empirical question with one of distributive justice. If HTEs are analysed sensibly, this does not have to be the case, but again, in the absence of a good critical literature, it is worth worrying that joint decision-making might stress the outputs of an algorithm over justice concerns \citep{Green2019Disparate}.

Treatment effects tell us nothing about justice, and if those worse off in a society have lower treatment effects, this should not mean they receive less. While it is common to assume that the benefits of programs are concave, that is those with less will benefit more from each marginal dollar \citep{Kaplow2003Concavity}, this is not always the case with regard to real-world deprivations and disadvantage. One particularly egregious example of this is in cases where one type of disadvantage can reduce the treatment effect of an intervention designed to address another. For example, a study in Niger showed that a transfer payment to extremely poor people improved outcomes more for those who received their payment electronically via a mobile money service rather than as physical cash (Niger is one of the most ‘unbanked’ countries in the world so bank transfers were not a feasible design) \citep{Aker2016Payment}. While the study only covered areas with mobile coverage (and provided phones to households that lacked one), one can imagine that had the mobile transfers rolled out more widely, those in more remote areas with poor or no mobile coverage would have not experienced these benefits. They would have either had to take time to collect cash or have had unreliable access to their money. In other words, the CATE given existing deprivation in mobile access would have been lower than the CATE given adequate mobile access ceteris paribus. A closely related issue is difficult-to-spot SUTVA violations which manifest as treatment effect heterogeneity as laid out by \citet{Heiler2022Effect}. They use the example of a vocational training program where treatment effect heterogeneity is actually explicable as a difference in treatment, with men being more likely to train in trades while women were more likely to train in lower paid service jobs. The upshot of this is that HTE does not even reflect actual treatment effects, let alone how much one deserves an intervention, even under a framework purely concerned with maximising total treatment effect of the treated from an intervention.

The problem here is obviously not in giving decision makers HTE information; in fact this is useful information in identifying where programs are failing those participating in them. Instead, the problem would arise from relying too much on these estimates and not enough on judgements around distributive justice due to automation bias, essentially shifting the power of the system in the direction of direct power on the margin. While HTE estimation can help to shed light on issues around the performance of programs for different marginalised groups, it is important that HTEs are not used as a basis to decide who is eligible for a program and who is not. There may of course be cases where individuals are given an alternative program instead in order to better serve their needs and this is philosophically different from just denying a group with low but positive HTEs access to a program on the grounds of cost. Ultimately, eligibility decisions should not be solely quantitative judgements – issues of justice need to play a role as well. It is important not to lose sight of this even if HTE and policy learning approaches may emphasise the former.

\section{Conclusion} \label{conclusion}
HTE estimating models pose fairness concerns, however these are not necessarily of the kind raised by the \textit{AI Fairness} literature. As models designed for the most part to inform decisions rather than make them, the fairness approach involved depends on how directly a system has power over decisions. In some cases where an HTE estimating model is given relatively direct power over policy, an \textit{AI Fairness} approach could be useful. However, in cases where the power of the model is more indirect, unfair outcomes tend to be the result of biases in the human decision-maker that are induced by epistemic concerns, but can ultimately be addressed by helping the human decision-maker to conduct better analysis and critically weigh algorithmic evidence. In particular, the novelty of these methods means there is a lack of experience with them and a lack of established best-practice that could exacerbate unfairness.

While this paper has attempted to begin a critical discussion of causal machine learning methods, it is not an exhaustive survey of possible harms and benefits. While HTE analysis with manually specified effects is commonly used in policy evaluation, this paper has contended that methods that flexibly estimate HTEs from high-dimensional data bring with them \textit{AI Fairness} concerns that make them qualitatively different. As these methods are yet to be widely used in practice for policy-making, we are limited to speculation in our ability to understand possible problems. However, there may be other harms that we have not addressed because they are beyond our speculation. It is also worth noting that our decision to defer to human decision-makers for value judgements prevented us from engaging with arguments that may be important for understanding the fairness implications in the use of HTE estimating models. For example, it may be that the availability of HTE estimating models actually encourages an informed decision by policy-makers to move from human-led decision making to empowering policy learners to make these decisions instead. In addition, it is possible that HTE estimating models in their complexity can allow for policy-makers to make decisions that are unpopular and in some broader sense unfair by using the inscrutability of the decision-making model to hide the decision, or the decision-making power of the system to avoid personal responsibility for decisions – a ‘hiding behind the computer effect’ \citep{Mittelstadt2016ethics,Zarsky2016Trouble}.

HTE estimation with causal ML methods is potentially a very powerful tool for policy evaluation, however it is an analytical method which brings with it drawbacks of a kind that do not exist in traditional methods. Given the stakes of policy decisions, it is important to understand the fairness implications of these methods before using them for decision-making.

\bibliographystyle{agsm}
\setcitestyle{authoryear,open={(},close={)}}
\bibliography{references}

\end{document}